\documentclass[letterpaper, 10 pt, conference]{ieeeconf}  

\IEEEoverridecommandlockouts                              

\usepackage{cite}

\title{Co-optimization of Speed and Gearshift Control for Battery Electric Vehicles Using Preview Information}

\author{Kyoungseok Han, Nan Li, Ilya Kolmanovsky, Anouck Girard, Yan Wang, Dimitar Filev, and Edward Dai
\thanks{*This work was supported by the Ford Motor Company}
\thanks{$^{1}$Kyoungseok Han, Nan Li, Ilya Kolmanovsky, and Anouck Girard are with the Department of Aerospace Engineering, University of Michigan, Ann Arbor, MI 48109, USA
        {\tt\small \{kyoungsh,nanli, ilya,anouck\}@umich.edu}}%
\thanks{$^{2}$Ford researchers are with the Ford Motor Company, Michigan, Dearborn, MI 48124, USA
        {\tt\small \{ywang21,dfilev,zdai\}@ford.com}}%
}

\usepackage{epsfig}
\usepackage{epstopdf}
\usepackage{subcaption}
\usepackage{multirow}
\usepackage{graphics} 
\usepackage{epsfig} 
\usepackage{amsmath} 
\usepackage{amssymb}  
\usepackage{bm}
\usepackage{epstopdf}
\usepackage{color}
\usepackage{xcolor}
\usepackage[linesnumbered]{algorithm2e}
\usepackage{enumerate}
\usepackage{footnote}
\usepackage{array}
\newcolumntype{P}[1]{>{\centering\arraybackslash}p{#1}}
\makesavenoteenv{tabular}

\begin{document}

\maketitle

\begin{abstract}
This paper addresses the co-optimization of speed and gearshift control for battery electric vehicles using short-range traffic information. To achieve greater electric motor efficiency, a multi-speed transmission is employed, whose control involves discrete-valued gearshift signals. To overcome the computational difficulties in solving the integrated speed-and-gearshift optimal control problem that involves both continuous and discrete-valued optimization variables, we propose a hierarchical procedure to decompose the integrated hybrid problem into purely continuous and discrete sub-problems, each of which can be efficiently solved. We show, by simulations in various driving scenarios, that the co-optimization of speed and gearshift control using our proposed hierarchical procedure can achieve greater energy efficiency than other typical approaches.
\end{abstract}

\IEEEpeerreviewmaketitle

\section{Introduction}
Over the past few years, there has been a growing research and commercial interest in the battery electric vehicles (BEVs) due to their potential of environmental benefits and low operating cost \cite{williamson2015industrial}. The control of connected and automated vehicles (CAVs) has also been of increasing interest to provide safer and more efficient transportation \cite{sun2018robust, han2018safe, vahidi2018energy}. Although both technologies have their own strengths, synergistic effects can be created when they are combined.

CAVs, in particular, are able to exploit real-time traffic information. For example, advanced driver-assistance systems (ADAS) with multiple data sources such as LiDAR, radar, and vehicle-to-vehicle (V2V) communications can provide and utilize short-range traffic information. In addition, advanced spatial localization technologies such as the global positioning system (GPS), geographic information system (GIS), and vehicle-to-infrastructure (V2I) communications enable the vehicles to obtain and utilize long-range information. In this paper, a control strategy for BEVs aimed at improving their energy efficiency and extending driving range is proposed. 

\subsection{Literature Review}
Optimizing vehicle speed while satisfying safety constraints has been extensively investigated in the literature. Dynamic Programming (DP) is often used to find the globally optimal speed trajectory, assuming that the entire route is given \cite{ozatay2014cloud, zeng2018globally}. Stochastic DP can partially alleviate the need for perfect information of the entire route \cite{mcdonough2014stochastic}. However, DP-based solutions may require heavy computations and, still, previews of the entire route. 

To keep online computations manageable, several researchers have proposed Pontryagin's Minimum Principle (PMP)-based speed planning strategies \cite{wan2016optimal, saerens2013calculation}, which require less computational effort compared to DP. Using PMP, an energy management strategy for hybrid electric vehicles including gearshift strategy has also been proposed in \cite{ngo2012optimal} based on a given speed profile. However, PMP only provides necessary conditions for optimality and the two-point boundary value problem associated with PMP conditions may not be easy to handle numerically. It is revealed in  \cite{asher2017prediction} that prediction errors of the future speed trajectory may have a significant impact on the energy consumption. Therefore, it may be better to update predictions in real time based on latest traffic information. In turn, control should adapt to the new predictions when new traffic information becomes available. Short-range preview information and Model Predictive Control (MPC) are used in \cite{prakash2016assessing} to control the vehicle speed while satisfying following-distance and speed constraints. Also, the approach proposed in \cite{seok2018energy} also optimizes the speed trajectory over a short horizon, i.e., 7 sec, using short-range preview information to minimize the cumulative wheel power consumption. Parametric optimization of a feedback-based vehicle speed controller to improve fuel economy is presented in \cite{li2016sequential}. It should also be noted that the optimized vehicle speed profiles are different depending on the powertrain types. For combustion engine, it is well known that pulse-and-glide speed profile gives benefit of energy consumption, but smoothed velocity works for BEVs \cite{han2019fundamentals}.

Another approach to improve energy efficiency of BEVs is the introduction of multi-speed transmissions. Traditionally, unlike in the powertrain of combustion engines, only a single reduction gear was used for BEVs \cite{dib2014optimal}. As a result, the operating point of the motor cannot be adjusted for given vehicle speed and acceleration command. However, in recent years, multi-speed transmissions for BEVs, i.e., 2-4 speed transmissions, have been considered \cite{guo2017line, pakniyat2014gear, han2018acc} to achieve further energy efficiency improvement. 

While extensive literature on the benefits of speed planning for CAVs and usage of multi-speed transmissions for BEVs exist, little research has been reported addressing the combination of these two technologies. In \cite{hu2017integrated}, optimizations at the vehicle level and at the powertrain level are performed concurrently for application to cruise control, in which the vehicle is controlled to track a constant reference speed and real-time traffic information is not exploited. Co-optimization of vehicle speed and powertrain control according to signal timing of traffic lights is presented in \cite{guo2017optimal}. 

Because of the discrete-valued gear ratios of a multi-speed transmission, the system representing the BEV's speed-and-powertrain coupled dynamics is essentially a hybrid system\footnote{More specifically, a system with mode switches.}. Exactly solving an optimal control problem for such a system usually involves solving a mixed integer nonlinear program, which is typically computationally demanding \cite{di2008control}. In \cite{guo2017optimal}, the speed trajectory is optimized first using a cost function that approximately represents the energy consumption and does not involve powertrain variables; the gear ratio is then selected for the given speed trajectory. When selecting the gear ratio, it is assumed that the gear ratio is set constant over the control horizon and, as a result, the gearshift optimal control involves only a single discrete-valued optimization variable. Although such a decomposition of speed planning and gear ratio selection as well as the constant gear ratio assumption simplify the computations, multiple gearshifts over the control horizon may lead to better fuel economy.

\subsection{Research Contributions}
To overcome the computational difficulty in exactly solving the speed-and-gearshift co-optimization problem as well as the above limitations, in this paper we propose a hierarchical procedure to decompose the integrated hybrid problem into purely continuous and discrete sub-problems. It will be shown that the sub-problems, and hence the original integrated problem, can be solved with manageable computational effort. We integrate such a hierarchical procedure in a model predictive control framework, using short-horizon preview information of traffic flow to improve BEV energy efficiency.


In particular, based on simulation results in various driving scenarios, we show that: (1) Smoothing vehicle speed can improve BEV energy efficiency, which is in line with the conclusions reached in \cite{prakash2016assessing, prakash2018role}. (2) The employment of multi-speed transmissions can further improve BEV energy efficiency, in particular, up to $19.67\%$ compared to using a single reduction gear. (3) Co-optimization of speed   and gearshift control by our proposed approach using short-horizon preview information can achieve greater energy efficiency than other conventional approaches, and closely matches the results where the gearshifts are optimized using DP with perfect preview information of the entire trip.

\subsection{Paper Organization}
The paper is organized as follows. In Section~II, the system model is described together with the problem statement. The hierarchical procedure to solve the hybrid speed-and-gearshift co-optimization problem is introduced in Section~III. The effectiveness of the proposed approach is illustrated in Section~IV. Finally, discussions and conclusions are given in Section~V.

\section{System Model And Problem Statement}

\subsection{System Model}
\subsubsection{Vehicle Dynamics}
We use the following model with state variables $x^v = [s, v]'$ to represent vehicle longitudinal dynamics:
\begin{align} 
    &\dot{s}=v, \label{system_model}\\ 
    &\dot{v}=\frac{T_w}{m r_{w}}-\frac{1}{2m}\rho A_{f} C_{d} v^2 - g\sin\theta - f g\cos\theta, \label{system_model2}
\end{align}
where $s$ is the vehicle travel distance, $v$ is the vehicle speed, 
$m$ is the vehicle weight\footnote{It is assumed that the vehicle's effective mass, which accounts for both static mass and rotational inertia effects, is approximately equal to the vehicle's mass.}, $T_{w}$ is the wheel torque, $r_{w}$ is the tire radius, $\rho$ is the air density, $A_{f}$ is the frontal area of the vehicle, $C_{d}$ is the aerodynamic drag coefficient, $\theta$ is the road inclination, $g$ is the gravitational constant, and $f$ is the coefficient of rolling resistance.

Control input $T_w$ is determined by the motor torque $T_m$, friction brake torque $T_b$, selected gear ratio $i_g$, and final drive ratio $i_0$ as follows:
\begin{equation}\label{wheel_torque}
    T_w = T_m\, i_g\, i_0 - T_b.
\end{equation}
Here, $T_b$ is applied when the minimum value of $T_m$, i.e., $T_{m}^{\min}$, is insufficient to generate the required wheel torque. To simplify the problem, we assume that motor torque is always sufficient to track the requested wheel torque without the need for friction braking, i.e., $T_w = T_m\, i_g\, i_0$.

Using Euler's method, the model for vehicle longitudinal dynamics is discretized as follows:
\begin{equation}\label{vehicle_model}
    x_{k+1}^{v} = x_k^v + f^v(x_k^v, u_k^v) \,T_s,
\end{equation}
where $x_k^v=[s_k, v_k]'$, $u_k^v = T_{w,k}$, $T_s = 1$~sec is the sampling period, $f^v(\cdot)$ is the nonlinear dynamic model (\ref{system_model}) and (\ref{system_model2}), and subscript $k\in \mathbb{Z}_{\ge 0}$ indicates the discrete time step.

Note that actual control inputs are the motor torque $T_m$ and gear ratio $i_g$, but these are merged into $T_w = T_m\, i_g\, i_0$ in this subsection.

\subsubsection{Battery Dynamics}
The battery state-of-charge (SoC) dynamics are described by:
\begin{equation}\label{SoC}
    \dot{SoC}(t)=-\frac{I_b(t)}{C}=-\frac{V_{oc}(t)-\sqrt{V_{oc}^2(t)-4R_b(t)P_b(t)}}{2\,C R_{b}(t)},
\end{equation}
where $I_b$ is the battery current, $C$ is the battery capacity, $V_{oc}$ is the open-circuit voltage in series with the battery resistance $R_{b}$, and $P_b$ is the consumed battery power.

The model for SoC dynamics is also discretized, as follows:
\begin{equation}\label{SoC_discretized}
    x^s_{k+1} = x^s_k + f^s(x_k^s, u_k^s) \,T_s,
\end{equation}
where $x_k^s=SoC_k$, $u_k^s = P_{b,k}$, and $f^s(\cdot)$ is the model (\ref{SoC}).

In many studies, $V_{oc} \text{ and } R_b$, which are functions of SoC, i.e., $V_{oc}(t) = V_{oc}\big(SoC(t)\big)$ and $R_b(t) = R_b\big(SoC(t)\big)$, are assumed to be constant \cite{kim2011optimal}. Such an assumption reduces the model complexity but also decreases the model fidelity. In this paper, to model the SoC dynamics more accurately, the variations of $V_{oc} \text{ and } R_b$ with SoC are accounted for in the prediction model.

The SoC value at the end of a planning horizon of length $N$, $SoC_{k+N|k}$, can be predicted as follows:
\begin{align}\label{SoC_predict}
& SoC_{k+N|k}=SoC_k - \nonumber \\
& \frac{T_s}{2\, C} \sum_{i=k}^{k+N-1}  \frac{V_{oc,i|k}-\sqrt{V_{oc,i|k}^2-4R_{b,i|k}P_{b,i|k}}}{R_{b,i|k}} \,, 
\end{align}
where $SoC_{k}$ represents the current SoC value.

Although (\ref{SoC_predict}) accounts for variations of $V_{oc} \text{ and } R_b$ with SoC, changes are relatively small. Therefore, battery powers $P_{b,i|k}$, $i=k,\dots,k+N-1$, are the major factors in determining $SoC_{k+N|k}$. It is expressed as:
\begin{equation}\label{Pb}
    P_{b,i|k} = \begin{cases} \frac{P_{m,i|k}}{\eta_{b}^+} = \frac{w_{m,i|k} T_{m,i|k}}{\eta_{b}^+\, \eta_{m,i|k}} & \text{for } T_{m,i|k} \geq 0, \\
    \frac{P_{m,i|k}}{\eta_{b}^-} = \frac{w_{m,i|k} T_{m,i|k}}{\eta_{b}^-\, \eta_{m,i|k}} & \text{for } T_{m,i|k} < 0, \end{cases}
\end{equation}
where $w_{m,i|k}$ is the motor speed, $T_{m,i|k}$ is the motor torque, $\eta_{b}^+ \in (0,1)$ is the battery-depletion efficiency, $\eta_{b}^- > 1$ is the battery-recharge efficiency, and $\eta_{m,i|k}$ is the motor efficiency determined by the motor operating point, i.e., $\eta_{m,i|k} = \eta_m (\omega_{m,i|k},T_{m,i|k})$.

Here, the predicted wheel torque $T_{w,i|k}$, vehicle speed $v_{i|k}$, and gear ratio $i_{g,i|k}$ determine the motor operating point, i.e., $\omega_{m,i|k}$ and $T_{m,i|k}$, as follows: 
\begin{align}
    \omega_{m,i|k} &= i_{g,i|k}\, i_0\, \omega_{w,i|k} = i_{g,i|k}\, i_0\, \frac{v_{i|k}}{r_w}, \label{wheel_spd}\\
    T_{m,i|k} &= \frac{T_{w,i|k}}{i_{g,i|k}\, i_0},
\end{align}
where $\omega_{w,i|k}$ is the predicted wheel rotational speed.

By \eqref{SoC} and \eqref{wheel_spd}, it can be said that minimization of current fluctuation minimizes the SoC consumption, which is equivalent to vehicle speed smoothing.
\subsection{Problem Statement}
In this subsection, the optimal control problem is formulated based on the models  for vehicle and battery dynamics developed in Section II-A. The discrete-time models \eqref{vehicle_model} and \eqref{SoC_discretized} are aggregated as:
\begin{equation}\label{combined_model}
    x_{k+1}^c = x_k^c + f^c(x_k^c,u_k^c) \,T_s,
\end{equation}
where the state variables are $x_k^c = [s_k, v_k, SoC_k]'$, the control inputs are $u_k^c = [T_{m,k}, i_{g,k}]'$, and 
\begin{equation} \label{augmented_model}
    f^c(x_k^c,u_k^c) = \begin{bmatrix}
f^v(x_k^v,u_k^v)\\  f^s(x_k^s,u_k^s)
\end{bmatrix}.
\end{equation}

The control objective is to minimize the battery SoC consumption by concurrently controlling the motor torque $T_{m}$ and gear ratio $i_{g}$ while keeping the vehicle speed $v$ within the acceptable range $[v^{\text{min}},v^{\text{max}}]$ and following a reference speed trajectory $v^r$ within acceptable tolerance. 

In this paper, we assume that a short-horizon preview of the reference vehicle speed, $v^r$, is available.  Such a preview can be generated by predicting the flow speed of the vehicles in front \cite{lefevre2014comparison}. The receding-horizon formulation of this optimal control problem is given by
\begin{subequations}\label{orig_opt}
\begin{align}
    \text{min}\quad & J_{k} =\sum_{i=k}^{k+N-1} L(x^c_{i+1|k}, u^c_{i|k}), \nonumber\\
            \text{s.t.}\quad & L(x^c_{i+1|k}, u^c_{i|k}) = -w_1\, \Delta SoC_{i+1|k} + \cdots \nonumber \\ 
                & + w_2\, (v_{i+1|k}-v_{i+1|k}^r)^2 + w_3\, \Delta T_{m,i|k}^2, \\
                & x_{i+1|k}^c=x_{i|k}^c+ f^c(x_{i|k}^c,u_{i|k}^c)\, T_s, \label{cont_o_s} \\
                & v^{\text{min}} \leq v_{i+1|k} \leq v^{\text{max}}, \\
                & \max(0.9v^{r}_{i|k}, 0.5) \leq v_{i|k} \leq 1.1v^{r}_{i|k}, \label{speed_const} \\
                & \tau_{\text{min}}(v_{i|k}+\delta)\leq s_{i|k}-s_{i|k}^r \leq \tau_{\text{max}}(v_{i|k}+\delta) \label{pos_const}\\
                -& T_{m}^{\text{max}} (w_{m,i|k}) \leq T_{m,i|k} \leq T_{m}^{\text{max}}(w_{m,i|k}),  \\
                & i_{g,i|k} = i_{g}(\eta_{g,i|k}), \label{geardy_1}  \\
                & \eta_{g,i+1|k} = \eta_{g,i|k} + u_{g,i|k}, \label{geardy_2} \\
                & \eta_{g,i|k} \in \{1,\cdots, \eta_g^{\max}\},  \label{geardy_3} \\
                & u_{g,i|k} \in U_g = \{-1, 0, 1\}, \label{geardy_4} \\
                & \qquad \qquad i=k,\cdots,k+N-1, \nonumber  
\end{align} 
\end{subequations}
 with respect to $u_{i|k} = [T_{m,i|k}, u_{g,i|k}]'$, $i=k,\cdots,k+N-1$, where $\Delta SoC_{i+1|k} = SoC_{i+1|k} - SoC_{i|k}$, $\Delta T_{m,i|k}=T_{m,i|k}-T_{m,i-1|k}$, $T_{m}^{\text{max}}(\cdot)$ is the maximum available motor torque depending on motor speed $w_{m}$, and $w_1, w_2, \text{ and } w_3$ are non-negative weighting factors.

In the formulated receding-horizon optimization problem \eqref{orig_opt}, the terms $(v_{i+1|k}-v_{i+1|k}^r)^2$ and $\Delta T_{m,i|k}^2$ in the cost function are used, respectively, for following the reference speed trajectory and for smoothing the speed trajectory and reducing battery current spikes (steep changes of battery current), which impede energy efficiency and battery lifespan. In (\ref{geardy_1}-\ref{geardy_4}), gear ratio dynamics are augmented to incorporate the requirement that no gear skipping is allowed in our gear system, where $\eta_{g,i|k} \in \{1,\cdots, \eta_g^{\max}\}$ indicates the gear positions and $u_{g,i|k} \in U_g = \{-1, 0, 1\}$ represents the gearshift signals. The elements $-1$ and $1$ of $U_g$ denote the down- and up- shift signals respectively, and 0 means the signal to maintain the gear position. In particular, a three-speed transmission is assumed to be employed, i.e., $\eta_{g,i|k} \in \left \{ 1, 2, 3 \right \}$, and the corresponding gear ratios are $i_g(\eta_{g})$.

It should be noted that the control inputs include both the continuous-valued signal $T_{m}$ and the discrete-valued signal $u_{g}$, which makes the system essentially a switching/hybrid system. As mentioned in Section~I, traditional approaches to solve such a hybrid optimization problem \eqref{orig_opt} involving mixed integer nonlinear programming can be computationally demanding \cite{liberzon2003switching, goebel2009hybrid}. In the next section, we propose a hierarchical approach to approximately solve \eqref{orig_opt} by decomposing it into multiple simpler sub-problems.

\begin{figure} [t]
\begin{center}
  \includegraphics[width=80mm]{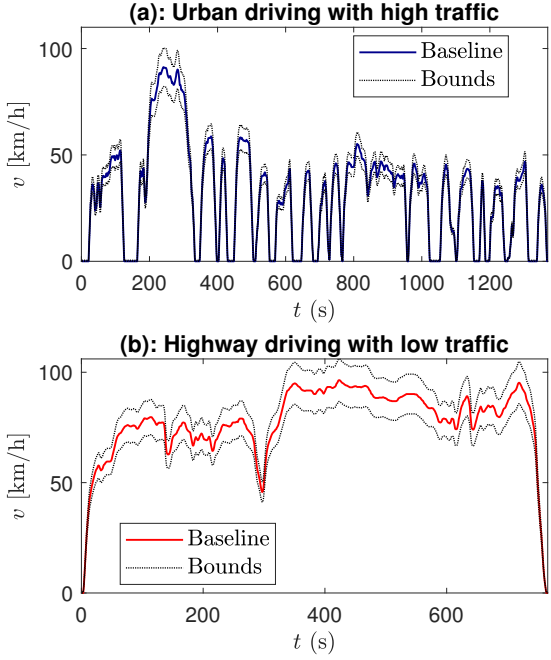}
  \caption{Reference speed profiles. (a) The speed trajectory of urban driving (blue) and its speed bounds (black dashed); (b) The speed trajectory of highway driving (red) and its speed bounds.}
  \label{drive_cycle}
\end{center}
\end{figure}

To avoid issues in recursive feasibility when repeatedly solving the optimization problem \eqref{orig_opt}, the constraints \eqref{speed_const} and \eqref{pos_const} are imposed as \textit{soft constraints}, i.e., through penalties for violations. In this paper, the reference speed profiles are based on standard driving cycles and are assumed to be the leader car's speed. Also, the target vehicle is controlled to keep an appropriate distance between itself and the leader car by \eqref{pos_const}. The omitted symbols in \eqref{orig_opt} can be found in Table.~\ref{model_pram}.

Fig.~\ref{drive_cycle} shows examples of reference speed profiles, where the dashed lines indicate the speed error bounds corresponding to \eqref{speed_const}. Fig.~\ref{drive_cycle}(a) shows the urban driving cycle and Fig.~\ref{drive_cycle}(b) shows the highway driving cycle. 

\section{Hierarchical Speed-and-Gearshift Control Co-Optimization Solution Method}
\begin{figure} [t]
\begin{center}
  \includegraphics[width=75mm]{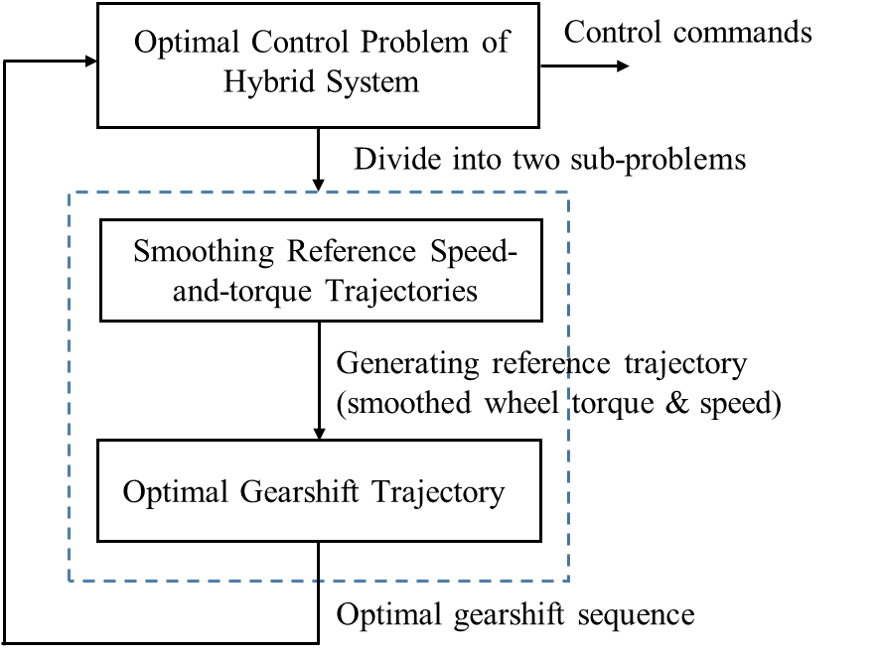}
  \caption{Hierarchical solution procedure.}
  \label{control_architecture}
\end{center}
\end{figure}
Fig.~\ref{control_architecture} illustrates the proposed hierarchical approach in approximately solving the hybrid optimization problem \eqref{orig_opt}. In particular, \eqref{orig_opt} is decomposed into three purely continuous or discrete sub-problems: In the $1^{st}$ sub-problem, the discrete-valued variables, i.e., the gearshift signal, are excluded and only the continuous-valued variables, i.e., the wheel torque and speed, are adjusted. As a result, the $1^{st}$ sub-problem is a standard nonlinear program. The solution to the $1^{st}$ sub-problem is used in the $2^{nd}$ sub-problem,  where the gearshift signal is decided. The $2^{nd}$ sub-problem is essentially a search problem of finding the minimum element within a small, finite set of numbers, which requires small computational effort. Finally, the gearshift signal is passed to the original problem \eqref{orig_opt} to eliminate discrete-valued unknowns and this results in a standard nonlinear program. Note that no iterations are involved and the three optimization problems are solved as a single pass.

\subsection{Smoothing Reference Speed-and-Torque Trajectories}

Motivated by the observations in \cite{ozatay2014cloud,zeng2018globally,wan2016optimal,prakash2016assessing,seok2018energy} and the physics-based analytic results in \cite{li2016sequential,he2018data} that fuel economy can be improved by reducing velocity fluctuations, we formulate the following problem to smooth the speed by attenuating torque changes:
\begin{subequations}\label{1st_subprob}
\begin{align}
     \text{min}\quad & J^{1}_k =\sum_{i=k}^{k+N-1} \Big( {w_2(v_{i+1|k}-v_{i+1|k}^r)^2+w_3\Delta T_{w,i|k}^2} \Big) \nonumber \\
            \text{s.t.}\quad & x_{i+1|k}^v=x_{i|k}^v+f^v(x_{i|k}^v,u_{i|k}^v)\, T_s, \\
            & v^{\text{min}} \leq v_{i+1|k} \leq v^{\text{max}}, \\
            & \max(0.9v^{r}_{i|k}, 0.5) \leq v_{i|k} \leq 1.1v^{r}_{i|k}, \\
            & \tau_{\text{min}}(v_{i|k}+\delta)\leq s_{i|k}-s_{i|k}^r \leq \tau_{\text{max}}(v_{i|k}+\delta),\\
            & |T_{w,i|k}| \leq T_{m}^{\text{max}} \Big(i_g^{\text{max}}\, i_0\, \frac{v_{i|k}}{r_w}\Big)\, i_g^{\text{max}}\, i_0, \\
            & \qquad \qquad i= k,\cdots,k+N-1, \nonumber  
\end{align} 
\end{subequations}
where $i_g^{\text{max}}$ is the maximum gear ratio.

The important features of this problem are that the SoC dynamics are not involved and the only optimization variables are the continuous-valued wheel torques $T_{w,i|k}$, $i=k,\cdots,k+N-1$.

The optimal sequence of wheel torques, $T_{w,i|k}^*$, is solved for and the corresponding vehicle speeds, $v_{i|k}^*$, are obtained. They are passed to the $2^{nd}$ sub-problem to find the optimal gearshift sequence as follows. 

\subsection{Optimal Gearshift Trajectory}

Let $\pi_k = \{u_{g,k|k}, u_{g,k+1|k},...,u_{g,k+N-1|k}\}$ be a gearshift sequence, taking values in $\Pi_k$, the set of all admissible gearshift sequences. The gearshift sequence is decided by solving the following optimization problem:
\begin{align} \label{gear_1}
     \underset{\pi_k \in \Pi_k}{\text{min}} &J^{2}_k = - \sum_{i=k}^{k+N-1} \eta_m (\omega_{m,i|k},T_{m,i|k}),
\end{align}
where the motor efficiency function, $\eta_m (\omega_{m,i|k},T_{m,i|k})$, is provided by a lookup table, and
\begin{subequations}
\begin{align}
    \omega_{m,i|k} &= i_{g,i|k}\, i_0\, \frac{v_{i|k}^*}{r_w}, \\
    T_{m,i|k} &= \frac{T_{w,i|k}^*}{i_{g,i|k}\, i_0},
\end{align}
\end{subequations}
subject to
\begin{subequations}
\begin{align}
    i_{g,i|k} &= i_{g}({\eta_{g,i|k}}), \\
    \eta_{g,i+1|k} &= \eta_{g,i|k} + u_{g,i|k}, \\
    \frac{\big|T_{w,i|k}^*\big|}{i_{g,i|k}\, i_0} &\le T_{m}^{\text{max}} \Big(i_{g,i|k}\, i_0\, \frac{v_{i|k}^*}{r_w}\Big), \\
    i &=k,\cdots,k+N-1. \nonumber  
\end{align}
\end{subequations}

To avoid overly frequent gearshifts, when constructing the admissible set of gearshift sequences $\Pi_k$, we restrict the maximum number of gearshifts, $n_u^{\max}$, i.e., enforce the constraint
\begin{equation} \label{gearshift_num}
\sum_{i=k}^{k+N-1} |u_{g,i|k}| \le n_u^{\max}.
\end{equation}

\begin{figure} [t]
\begin{center}
  \includegraphics[width=75mm]{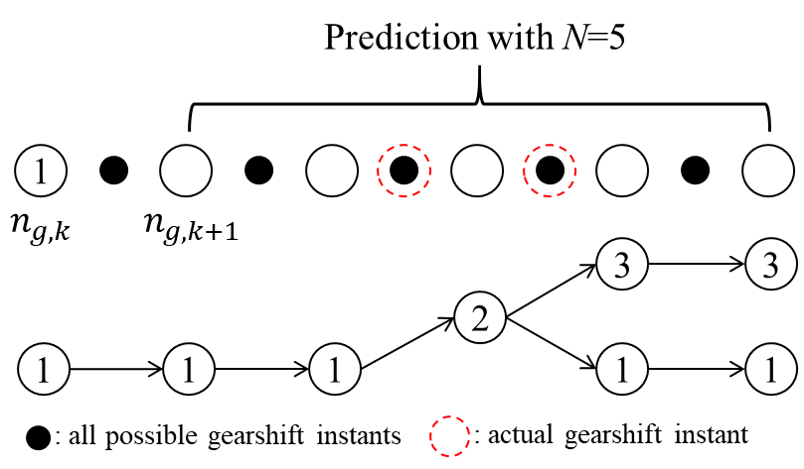}
  \caption{Examples of admissible gearshift profile.}
  \label{gearshift_description}
\end{center}
\end{figure}

Furthermore, the constraints $\eta_{g,i|k} \in \{1,\cdots, \eta_g^{\max}\}$ are also incorporated when constructing $\Pi_k$. Indeed, the set $\Pi_k$ satisfying the above requirements only depends on the current gear position $\eta_{g,k|k} \in \{1,2,3\}$. Thus, $\Pi_k = \Pi(\eta_{g,k|k}) \in \{\Pi^1,\Pi^2,\Pi^3\}$, where $\Pi^j$, $j=1,2,3$, can be pre-constructed offline, which gives computational benefit. See the example of admissible gearshift profile in Fig.~\ref{gearshift_description} when the current gear position is 1 and the gear change is allowed up to twice.

\subsection{Integrated Speed and Gearshift Control}

One important feature of our approach as compared to previous approaches, e.g., the ones in \cite{seok2018energy,guo2017optimal}, is that the smoothed speed and torque trajectories $(v_{i|k}^*,T_{w,i|k}^*)$ are not used as the final solution but only serve as references for deciding the gearshift sequence $u_{g,i|k}^*$.

In particular, after the wheel torques $T_{w,i|k}^*$ are obtained based on the optimization problem in Section III.A and the gearshift signals $u_{g,i|k}^*$ and the corresponding gear ratios $i_{g,i|k}^*$ are obtained based on the optimization problem in Section III.B, $u_{i|k}^* = [T_{m,i|k}^*, u_{g,i|k}^*]'$, $i=k,\cdots,k+N-1$, where $T_{m,i|k}^* = \frac{T_{w,i|k}^*}{i_{g,i|k}^* i_0}$, constitutes a feasible, sub-optimal solution to the original problem (\ref{orig_opt}). 
However, a better solution may be obtained by substituting $\{u_{g,k|k}^*, u_{g,k+1|k}^*,...,u_{g,k+N-1|k}^*\}$ into (\ref{orig_opt}) to eliminate the discrete-valued unknowns $u_{g,i|k}$, $\eta_{g,i|k}$, and $i_{g,i|k}$, $i=k,\cdots,k+N-1$, so that (\ref{orig_opt}) becomes a standard nonlinear program with $T_{m,i|k}$, $i=k,\cdots,k+N-1$, as the continuous-valued optimization variables. Furthermore, $T_{m,i|k}^*$, as a feasible, sub-optimal solution, is provided as the initial guess to the nonlinear programming solver, so that the solver iterates can converge to a local minimizer fast. We denote the obtained solution after solver convergence by $T_{m,i|k}^{**}$, and $u_{i|k}^{**} = [T_{m,i|k}^{**}, u_{g,i|k}^*]'$ is the final solution to (\ref{orig_opt}) by  our proposed approach. Note that the solution $u_{i|k}^{**}$ cannot be worse than the feasible initial guess $u_{i|k}^*$ in terms of cost values, meaning that this additional step to update $T_{m,i|k}^*$ to $T_{m,i|k}^{**}$ can always provide us with improved solutions.

After $u_{i|k}^{**}$, $i=k,\cdots,k+N-1$, is obtained, the first element $u_{k|k}^{**}$ is applied to the vehicle over one time step to update the vehicle's states $s,v,SoC$ as well as the gear $\eta$. The optimization problem (\ref{orig_opt}) with $k \rightarrow k+1$ is then solved again using the same hierarchical approach.

\section{Simulation Results}
In this section, we illustrate the effectiveness of the proposed approach on various driving cycles. The model parameter values are summarized in Table~\ref{model_pram}, which come from ADVISOR (a high-fidelity simulator for vehicle energy consumption analysis) \cite{wipke1999advisor}. 

Firstly, to verify our hypothesis that BEV energy efficiency can be improved by smoothing speed profiles, we apply the optimal solutions of the $1^{st}$ sub-problem to a BEV with a single reduction gear (without using a multi-speed transmission and thus without requiring gearshift control). After that, we apply the solutions of (\ref{orig_opt}) obtained using our proposed hierarchical procedure to the same BEV but equipped with a three-speed transmission to reveal the overall benefits of speed-and-gearshift control co-optimization. 

\begin{table}[!t]
\centering
\caption{Model Parameters} \label{model_pram}
\begin{tabular}{|c|c|c|}
\hline
Symbol            & Description                       & Value {[}Unit{]}                  \\ \hline
$m$               & Vehicle total mass                & 1445 {[}kg{]}                     \\ \hline
$r_w$             & Wheel radius                      & 0.3166 {[}m{]}                    \\ \hline
$A_f$             & Frontal area                      & 2.06 {[}$\text{m}^{2}${]}   \\ \hline
$C_d$             & Aerodynamic drag coefficient      & 0.312 {[}-{]}                     \\ \hline
$\rho$            & Air density                       & 1.2 {[}$\text{kg/m}^{3}${]} \\ \hline
$f$               & Coefficient of rolling resistance & 0.0086 {[}-{]}                    \\ \hline
$i_g(\eta_g)$ & Gear ratios, $\eta_{g} \in \left \{ 1, 2, 3 \right \}$         & \{3.05, 1.72, 0.92\}   \\ \hline
$i_0$ & Final drive ratio          & 4.2 {[}-{]}              \\ \hline
$i_g^s$           & Gear ratio of single gear    & 7.2 {[}-{]}             \\ \hline
$v^{\text{min}}, v^{\text{max}}$      & Acceptable range of vehicle speed   & \{0, 120\} {[}km/h{]}                       \\ \hline
$C$               & Battery capacity                  & 55 {[}Ah{]}                       \\ \hline
$\eta_b^+$        & Battery-depletion efficiency      & 0.9 {[}-{]}                       \\ \hline
$\eta_b^-$        & Battery-recharge efficiency       & 1.11 {[}-{]}                       \\ \hline
$n_u^{max}$        & Maximum number of gearshifts     & 1 {[}-{]}                       \\ \hline
$\tau_{\text{max}}, \tau_{\text{min}}$       & Maximum \& minimum headway     & \{1, 2\} {[}s{]}                       \\ \hline
$\delta$           & Minimum relative speed bound     & 5 {[}m/s{]}                       \\ \hline
$w_1, w_2, w_3$    & Weighting factors                & \{2000, 1, 1\}                  \\ \hline
\end{tabular}
\end{table}
\subsection{Pure Benefits from Smoothing Speed Profiles and Reducing Battery Current Spikes}
\begin{figure} [!t]
\begin{center}
  \includegraphics[width=90mm]{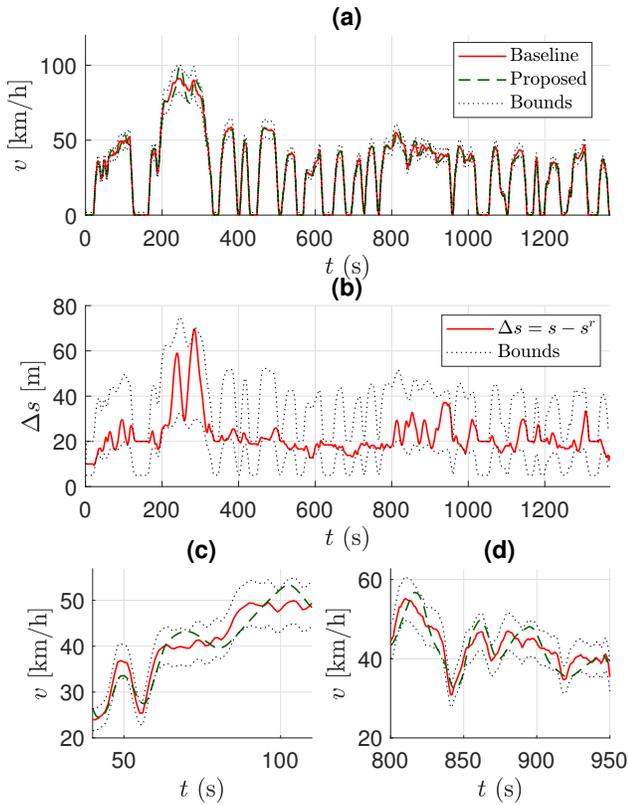}
  \caption{Simulation results on urban driving cycle (UDDS) with control based on speed smoothing. (a) The entire speed trajectories of original driving cycle (red), smoothed speed trajectory (green dashed), and speed bounds (black dotted); (b) The  trajectory of distance between the ego car and the leader car (red), and position bounds (black dotted); (c) and (d) The magnified speed trajectories in a particular area of interest.}
  \label{simul_1st}
\end{center}
\end{figure}

To verify the pure benefits from smoothing speed profiles and reducing battery current spikes, we apply the optimal solutions of the $1^{st}$ sub-problem to a BEV with a single reduction gear of ratio $i_g^s$. The motor torques are determined as follows:
\begin{equation}\label{motor_sequence}
    T_{m,i|k}^* = \frac{T_{w,i|k}^*}{i_g^s},\quad i=k, \cdots k+N-1. 
\end{equation}

Simulations are performed for four driving cycles with prediction horizon $N=5$, corresponding to preview of $5 T_s = 5$~sec. One of the detailed results is illustrated in Fig. \ref{simul_1st}. The red solid lines represent the trajectory of following the reference speed profile exactly (referred to as ``baseline''). The green dashed lines represent the trajectories of using the optimized control based on the solutions of the $1^{st}$ sub-problem and \eqref{motor_sequence} (referred to as ``speed smoothing with single reduction gear''). The black dotted lines represent the specified speed bounds. We can observe that the BEV can follow the reference speed profile within specified speed bounds while satisfying the position constraint Fig.~\ref{simul_1st}(b).

Looking at the trajectories more closely in Fig.~\ref{simul_1st}(c)-(d), we can observe that the vehicle speed trajectories of the BEV controlled based on solutions of the $1^{st}$ sub-problem are smoother -- high-frequency speed fluctuations are filtered out -- compared to those of following the reference speed profile exactly.

All simulations are conducted with the initial SoC=0.8, and the SoC consumption ($\Delta$SoC) results for the four driving cycles are summarized in Table~\ref{simul_2}. Although the performance in $\Delta \text{SoC}$ improvement of ``speed smoothing with single reduction gear'' compared to ``baseline'' depends on driving cycles, we note that $2.56\% \sim 10.42\%$ improvement can be achieved. This confirms the potential of smoothing vehicle speed, the strategy implemented based on the $1^{st}$ sub-problem, in improving BEV energy efficiency. 

\begin{table*}[t] 
\centering
\captionsetup{justification=centering}
\caption{Simulation results for different driving cycles. The improvement is reported over the Baseline (single gear, nominal speed profile).}\label{simul_2}
\begin{tabular}{|P{1.5cm}|P{1.4cm}|P{1.6cm}|P{1.6cm}|P{1.6cm}|P{1.85cm}|P{1.85cm}|P{1.85cm}|}
\hline
\multirow{2}{*}{}         & Baseline                                                   & \multicolumn{3}{c|}{Speed smoothing with single reduction gear}                                                                                                                        & \multicolumn{3}{c|}{Speed-and-gearshift co-optimization with 3-speed gears}                                                                                                     \\ \cline{2-8} 
                          & \begin{tabular}[c]{@{}c@{}}$\Delta$SoC \\ (\%)\end{tabular} & \begin{tabular}[c]{@{}c@{}}Prediction\\ Horizon\end{tabular} & \begin{tabular}[c]{@{}c@{}}$\Delta$SoC\\ (\%)\end{tabular} & \begin{tabular}[c]{@{}c@{}}Improvement\\ (\%)\end{tabular} & \begin{tabular}[c]{@{}c@{}}Prediction\\ Horizon\end{tabular} & \begin{tabular}[c]{@{}c@{}}$\Delta$SoC\\ (\%)\end{tabular} & \begin{tabular}[c]{@{}c@{}}Improvement\\ (\%)\end{tabular} \\ \hline
\multirow{2}{*}{UDDS}     & \multirow{2}{*}{7.47}                                      & $N=5$                                                        & 7.07                                                       & 5.35                                                       & $N=5$                                                        & 6.33                                                       & 14.69                                                      \\ \cline{3-8} 
                          &                                                            & $N=8$                                                        & 6.78                                                       & 9.24                                                       & $N=8$                                                        & 6.19                                                       & 16.57                                                      \\ \hline
\multirow{2}{*}{WLTC}     & \multirow{2}{*}{17.55}                                     & $N=5$                                                        & 17.1                                                       & 2.56                                                       & $N=5$                                                        & 16.12                                                      & 8.15                                                       \\ \cline{3-8} 
                          &                                                            & $N=8$                                                        & 16.3                                                       & 7.12                                                       & $N=8$                                                        & 15.84                                                      & 9.74                                                       \\ \hline
\multirow{2}{*}{LA92}     & \multirow{2}{*}{12.86}                                     & $N=5$                                                        & 11.79                                                      & 8.32                                                       & $N=5$                                                        & 10.44                                                      & 18.82                                                      \\ \cline{3-8} 
                          &                                                            & $N=8$                                                        & 11.52                                                      & 10.42                                                      & $N=8$                                                        & 10.33                                                      & 19.67                                                      \\ \hline
\multirow{2}{*}{US06 HWY} & \multirow{2}{*}{9.43}                                      & $N=5$                                                        & 9.01                                                       & 4.45                                                       & $N=5$                                                        & 8.64                                                       & 8.38                                                       \\ \cline{3-8} 
                          &                                                            & $N=8$                                                        & 8.91                                                       & 5.51                                                       & $N=8$                                                        & 8.61                                                       & 8.70                                                       \\ \hline
\end{tabular}
\end{table*}

\subsection{Overall Benefits from Speed-and-Gearshift Control Co-optimization}

\begin{figure} [!t]
\begin{center}
  \includegraphics[width=90mm]{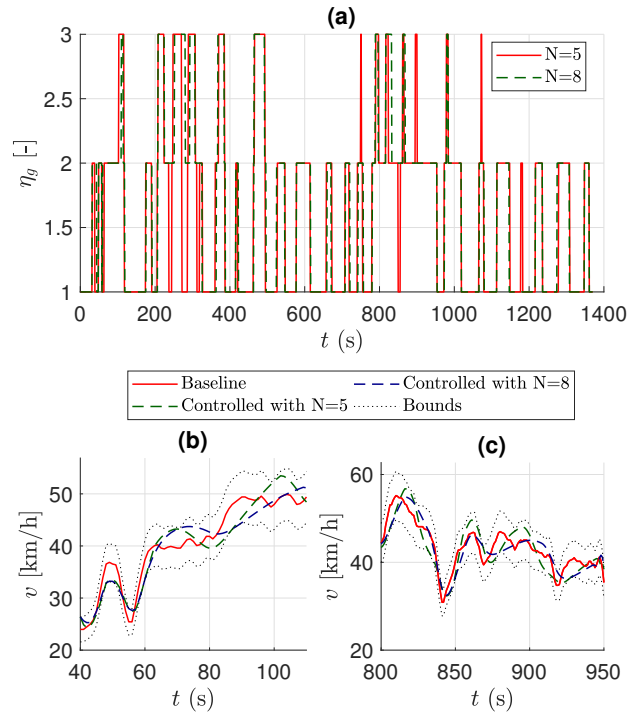}
  \caption{Simulation results on urban driving cycles (UDDS) with the integrated controls. (a): The entire gearshift trajectories with prediction horizons $N$=5 (red) and $N$=8 (green dashed); (b) and (c): The magnified speed trajectories on a particular area of interest: original driving cycle (red), smoothed speeds with $N$=5 (green dashed) and $N$=8 (blue dashed), and speed bounds (black dotted).}
  \label{simul_2nd}
\end{center}
\end{figure}

We now examine the overall benefits from speed-and-gearshift control co-optimization.
In particular, we consider a BEV with a three-speed transmission and apply the control based on the co-optimization problem \eqref{orig_opt} solved using our proposed hierarchical procedure. The obtained trajectories are referred to as ``speed-and-gearshift co-optimization with 3-speed transmission.''

The gear position profiles over the UDDS driving cycle are displayed in Fig.~\ref{simul_2nd}(a) for prediction horizons $N=5$ and $N=8$, which is a reasonable horizon as reported in the \cite{liu2019vehicle, seok2018energy}. We can observe that they are identical for most of the time and differ at a few places. Frequent gearshifts occur rarely, verifying the effectiveness of the constraint \eqref{gearshift_num}. In addition, we can see in Figs.~\ref{simul_2nd}(b) and (c) that the speed trajectory corresponding to a longer horizon $N=8$ is even smoother than that corresponding to a shorter horizon $N=5$, reflecting the effect of preview information.

\begin{figure} [!t]
\begin{center}
  \includegraphics[width=80mm]{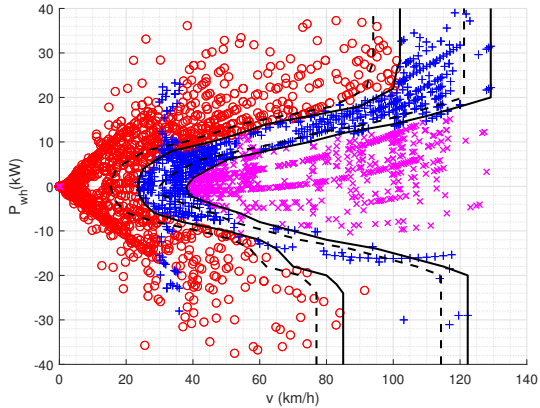}
  \caption{The gearshift schedule optimized using the approach in \cite{han2018acc}. Red 'o', blue '+', and pink 'x' correspond to operating points of the motor with the $1^{st}$ gear, $2^{nd}$ gear, and $3^{rd}$ gear, respectively.}
  \label{shift_map}
\end{center}
\end{figure}

The $\Delta$SoC results for the four driving cycles are summarized in Table~\ref{simul_2}. Compared to the results of using a single reduction gear, the use of a three-speed transmission and the speed-and-gearshift co-optimization can further improve BEV energy efficiency ($8.15\% \sim 19.67\%$ improvement for $N=5$, depending on the drive cycles).

In the above simulations, the average computation time to solve all of the three sub-problems in our hierarchical procedure for one step is $0.31$~sec for $N = 5$ and $0.51$~sec for $N = 8$. The computations are performed on the MATLAB R2018a platform running on an Intel Xeon E3-1246 3.50-GHz PC with 16.0GB of RAM. In particular, the nonlinear programs (NLPs) in the $1^{st}$ and $3^{rd}$ sub-problems (Sections III.A and III.C) are solved using the MATLAB $fmincon()$ function with the interior-point method and the computation times are determined using the MATLAB $tic-toc$ command. This implies the possibility of hardware realization of the control algorithm in real time.
Note that further reduction in computation times is possible by replacing $fmincon()$ with tailored NLP solvers, by exploiting inexact and real-time iteration solution strategies, and by symbolic and software optimization techniques, see e.g., \cite{walker2016design}.


Furthermore, to illustrate that co-optimization of speed and gearshift control using our proposed approach can achieve greater energy efficiency than optimizing speed and gearshift separately, we run two additional sets of simulations. Firstly,
we run simulations where the vehicle speeds are optimized based on the solutions of the $1^{st}$ sub-problem and the gearshifts are determined by a pre-defined shift schedule in Fig.~\ref{shift_map}. We note that the shift schedule has been optimized according to the driving cycles using the approach in \cite{han2018acc}. Secondly, we run simulations where the vehicle speeds are again optimized based on the $1^{st}$ sub-problem and the gearshifts are optimized using DP with the knowledge of the entire optimized speed profile. 
The $\Delta \text{SoC}$ results for the four driving cycles are summarized in Table~\ref{simul_3}. It can be observed that our approach outperforms the first approach (speed optimization + shift map) and closely matches the results of the DP (speed optimization + globally optimized gear shift trajectories).

\begin{table*}[t!] 
\centering
\captionsetup{justification=centering}
\caption{Results comparing speed-and-gearshift co-optimization and speed-and-gearshift separate-optimizations. }\label{simul_3}
\begin{tabular}{|P{1.0cm}|P{1.75cm}|P{1.75cm}|P{1.75cm}|P{1.75cm}|P{1.75cm}|P{1.75cm}|}
\hline
\multirow{2}{*}{}                                                    & \multicolumn{2}{c|}{\begin{tabular}[c]{@{}c@{}} Separate-optimization \\ with predefined gearshift map \end{tabular}}                 & \multicolumn{2}{c|}{\begin{tabular}[c]{@{}c@{}} Separate-optimization \\ with gearshift optimized using DP \end{tabular}}                     & \multicolumn{2}{c|}{\begin{tabular}[c]{@{}c@{}}Speed-and-gearshift \\ co-optimization \end{tabular}} \\ \cline{2-7} 
                                                                     & \begin{tabular}[c]{@{}c@{}}Prediction \\ horizon\end{tabular} & \begin{tabular}[c]{@{}c@{}}$\Delta$SoC \\ ($\%$)\end{tabular} & \begin{tabular}[c]{@{}c@{}}Prediction\\ horizon\end{tabular} & \begin{tabular}[c]{@{}c@{}}$\Delta$SoC\\ ($\%$)\end{tabular} & \begin{tabular}[c]{@{}c@{}}Prediction \\ horizon\end{tabular}   & \begin{tabular}[c]{@{}c@{}}$\Delta$SoC\\ ($\%$)\end{tabular}  \\ \hline
\multirow{2}{*}{UDDS}                                                & $N=5$                                                         & 6.83                                                           & $N=5$                                                        & 6.09                                                          & $N=5$                                                           & 6.33                                                       \\ \cline{2-7} 
                                                                     & $N=8$                                                        & 6.55                                                           & $N=8$                                                       &  5.79                                                         & $N=8$                                                          & 6.19                                                           \\ \hline
\multirow{2}{*}{WLTC}                                                & $N=5$                                                         & 16.72                                                          & $N=5$                                                        & 15.77                                                         & $N=5$                                                           & 16.12                                                         \\ \cline{2-7} 
                                                                     & $N=8$                                                        & 16.01                                                          & $N=8$                                                       & 15.28                                                         & $N=8$                                                          & 15.84                                                          \\ \hline
\multirow{2}{*}{LA92}                                                & $N=5$                                                         & 11.43                                                          & $N=5$                                                        & 10.15                                                         & $N=5$                                                           & 10.44                                                          \\ \cline{2-7} 
                                                                     & $N=8$                                                        & 11.17                                                          & $N=8$                                                       & 10.03                                                         & $N=8$                                                          & 10.33                                                          \\ \hline
\multirow{2}{*}{\begin{tabular}[c]{@{}c@{}}US06 \\ HWY\end{tabular}} & $N=5$                                                         & 8.88                                                           & $N=5$                                                        & 8.61                                                       & $N=5$                                                           & 8.64                                                          \\ \cline{2-7} 
                                                                     & $N=8$                                                        & 8.78                                                           & $N=8$                                                       & 8.56                                                          & $N=8$                                                          & 8.61                                                           \\ \hline
\end{tabular}
\end{table*}

\section{Conclusions}

This paper presents a strategy for battery electric vehicle speed-and-gearshift control co-optimization using preview information of traffic. The strategy optimizes continuous-valued vehicle speed and discrete-valued gearshift signal to minimize the energy consumption of BEVs equipped with multi-speed transmissions. By exploiting a hierarchical approach, we decompose the hybrid co-optimization problem of mixed integer nonlinear programming (MINLP) type into sub-problems with purely continuous or discrete-valued variables.
The simulation results suggest that substantial improvements in energy efficiency can be achieved by the proposed approach (up to 19.67$\%$) for different driving cycles. 
The proposed hierarchical approach reduces the computational cost in solving the hybrid co-optimization problem to a level that is comparable to the time available for real-time  implementations.


\bibliographystyle{IEEEtran}
\bibliography{ref}

\end{document}